\documentstyle[aps,twocolumn,prl,epsf,floats]{revtex}
\draft
\begin{document}
\title{On micro-bead mechanics with actin filaments}
\author{A. C. Maggs}
\address{PCT, ESPCI, 10 rue Vauquelin, 75231 Paris Cedex 05, France. }
\date{\today}
\maketitle 
\begin{abstract}
  Experiments have been performed using microscopic beads to probe the
  small scale mechanics of actin solutions. We show that that there
  are a number of regimes possible as a function of the size of the
  probing particle. In certain cases we argue 
  that the quasi-static response resembles a
  smectic crystal rather than an isotropic solid, implying
  an anomalous scaling of the mechanical response of actin
  solutions as a function of the size of the probing particles.
\end{abstract}
\pacs{87.15-v,83.10.Nm} The mechanics and rheology of actin filaments
are a beautiful model system for the study of the dynamics and
mechanics of semi-dilute polymers \cite{kas,kas2}. They are characterized
by length scales which are easily accessible with optical techniques
allowing the detailed study of phenomena such as tube dynamics.
However, the macroscopic rheology of these systems has been hard to
master from the experimental point of view. Difficulties of
purification and sample preparation lead to orders of magnitude
variations in such fundamental objects such as the value of the
plateau modulus
\cite{pollard,zaner,sackmanntube,sackmanntube2,janmey}, the standard
measure of the response of an entangled polymer solution to external
perturbations.

To get around these problems of macroscopic sample preparation and
also to probe the local viscoelastic behaviour of these materials a
number of experimental groups have started using small, colloidal
beads to study the local mechanics of these materials
\cite{zaner2,sackmannbead,amblard,schmidt1,fred,weitz}.  One either
pulls on the particles using super-paramagnetic beads in a magnetic
field, or one simply observes the fluctuations of the particles
undergoing Brownian motion.  In this letter I shall try to attack the
problem as to what exactly one measures in these experiments.  In
particular how large do these particles have to be in order to measure
a macroscopic elastic modulus and when do we expect to be sensitive to
the individual filament properties?  

In contrast with flexible polymers solutions there are two principal
length scales present in a semi-dilute solution of actin: the mesh
size and the persistence length. Naive application of scaling ideas
thus becomes a highly ambiguous exercise because an arbitrarily large
number of intermediated lengths can be created by considering
$\xi^{1-\alpha} l_p^{\alpha}$ with $\xi$ the mesh size and $l_p$ the
persistence length.  This ambiguity in lengths also translates into an
ambiguity in the plateau modulus which can be expressed as $k_B T$ per
characteristic volume.  As an example of this difficulty we might
quote two attempts to calculate the modulus in actin solutions
with scaling approaches \cite{fredmod,kroy} where two completely
different results are found due in part due to this problem. Indeed this
proliferation of lengths is already known for the tube geometry where
one finds both $\alpha =+1/5$ and $\alpha = -1/5$,
\cite{semenov,semenovbis,odjik}.  We shall show in this article that a
new intermediate scale with $\alpha=3/5$ becomes crucial in the
understanding of the elasticity of actin solutions at length scales
probed with micrometer sized beads. At these scales we show that the
elastic response is highly anisotropic and resembles that of a
smectic, with anomalous penetration of the response into the sample
and unusual scaling of the response with the size of the probing
particle.

Note that in this letter I am interested in the low frequency
mechanics and thus I exclude from the discussion high frequency
fluctuation measurements (up to $20KHz$) which have been recently
performed \cite{fred} and am interested in a quasi-static regime
between $10^{-3}Hz$ and $10Hz$.  The reason for considering this range
scale will become clear during the discussion.

A coherent picture of the large scale mechanics of non-crosslinked
actin solutions is now available.  The actin system is usually
polymerized \cite{schmidt1} in conditions such that the mean distance
between filaments, $\xi$ is between $0.3\mu$ and $1\mu$. $\xi$ can be
linked with the concentration of monomers $c$ by noting that
$\xi\sim1/\sqrt{c d}$ with $d$ the size of actin monomers. A useful
geometric quantity is the length of filament per unit volume $\rho
\sim 1/\xi^2$. The filament is characterized by its persistence length
$l_p$ which is close to $15\mu$ \cite{gittes}.  For a single weakly
bent filaments the energy of a configuration is given by \cite{landau}
\begin{equation} E = k_B T l_p/2 \int (\partial_s^2 {\bf r_{\perp}}(s))^2
  ds \label{landau} \end{equation} where \( {\bf r_{\perp}}(s)\) is
the transverse fluctuation of the filament about its equilibrium
shape.

In a manner which is familiar from flexible polymers the individual
filaments are confined to a tube whose diameter scales as $\xi^{6/5} /
l_p^{1/5}$ and the filament is confined to the tube by collisions
between the filament and its neighbors every $l_e \sim \xi^{4/5} l_p
^{1/5}$ \cite{semenov,semenovbis}. $l_e$ is is in some ways equivalent
to the entanglement length in the Doi-Edwards tube model
\cite{doi,degennes,morse}. The long time dynamics and mechanics are
dominated by the reptation of filaments along their tubes
\cite{sackmanntube,morse,me}.  This process has a characteristic time,
the reptation time, which defines the time scale beyond which the
sample behaves like a viscous fluid (rather than an elastic solid) and
can be as long as several hours \cite{sackmanntube}. Under macroscopic
shear the longitudinal stresses in a filament relax relatively rapidly
\cite{me} leaving a residual contribution to the free energy which
comes from the modification of the free energy of confinement of the
filament in its tube. A simple argument for this free energy is to
count $k_B T$ per collision of the tube with the filament. Thus the
macroscopic modulus varies as $G\sim\rho k_BT/l_e \sim
c^{1.4}/l_p^{1/5}$ as confirmed by an explicit calculation
\cite{morse,herve}.

This picture of filaments confined to a tube is only true on time
scales that are long enough for the filament to dynamically sample
fluctuations on the scale of $l_e$.  This time, which is determined by
the bending elasticity of the filaments varies as $\tau_e \sim \eta
l_e^4/l_p k_B T \sim 10Hz$ \cite{herve}. This is our reason for
restricting our treatment to lower frequencies, at higher frequencies
one is presumably sensitive to individual filament dynamics (coupled
by hydrodynamics) rather than the collective, entangled, modes that
interest us in this letter. For frequencies lower than the inverse
reptation time (ie frequencies comparable to $10^{-3}Hz$) the sample
behaves as a fluid and the bead moves freely as filaments slide out of
the way of the particles.

Before passing to the problem of the behavior of actin solutions we
shall revise a Peierls like argument from which we can deduce the basic
scaling behaviour of a normal elastic solid. We shall then adapt this
argument to the case of semiflexible filaments: Consider a bead of
radius $R$ embedded in an elastic medium in $d$ dimensions. If we pull
on the particle with a force $f$ we can make the following variational
ansatz in order to find the minimum energy configuration.
Let us assume that the material is disturbed over a distance $l$ from
the bead then the elastic energy, will scale in the following manner
\begin{equation}
E_{var} \sim G \int (\nabla a)^2 dV
\label{evar}
\end{equation}
where $a$ is an amplitude of displacement, $G$ an elastic constant and
the integral is over the variational volume $V \sim l^d$. This scales as 
\begin{equation} 
\label{evar2}
E_{var} \sim G (a/l)^2 l^d 
\end{equation}
We see that in less than two dimension an arbitrarily small force is
able to displace the bead large distances because $E_{var}$ can be
made small by increasing the variational parameter $l$.  In three
dimensions, however, the energy diverges with $l$ and has a lower bound for
small $l$ due to the short wavelength cutoff coming from the finite size of
the bead. Thus the minimum energy is found for $l \sim R$ and we
deduce that $E_{var} \approx G a^2 R$. At constant force the displacement scales
inversely with the bead size,
\begin{equation}
\label{inverse}
 a \approx  f/ G R
\end{equation}
A full calculation of the response of an
isotropic viscoelastic material has recently been performed and
confirms this simple scaling argument \cite{fred}.

We see that there is an anisotropy in the problem coming from the
direction in which  we apply the force $f$, and we should worry that the
volume excited is not spherical as has been assumed in the argument.
Let us perform a slightly more elaborate variational treatment where we
assume that the volume $V$ is characterized by an disk of
dimensions $l \times l \times D$ where the particle excites modes of
wavelength $l$ which penetrate $D$ into the sample in the direction of
$f$. In this case our estimate for $E_{var}$ is
\begin{equation}
\label{evar3}
E_{var} \sim (l^2 D) G ( ({{a}/ {l}} )^2 + ({{a}/ {D}})^2)
\end{equation}
Where $ a/l$ and $a/D$ are the estimates of the components of the
strain tensor in the material.  Taking $D$ as a
variational parameter one sees that $D \sim l$ and the problem
reduces to that considered above. This is in fact a crude statement of
the principle of St.  Vernet  that a force on a body with a
wavelength $l$ decays into the body over the same length scale, which
is a elementary property of periodic harmonic functions in three
dimensions.

How must this argument be modified in the actin system? Experiments
are performed with bead which vary in size from $.1 \mu $ and $10
\mu$.  The smallest beads pass between the filaments and diffuse
almost freely \cite{schmidt1}; they will not concern us any further.
Are we able to use continuum elastic arguments (like that above) to
deduce the experimental stress stain relationships?
We now argue that in actin solutions there are now two contributions
to the variational energy $E_{var}$. For large beads the normal
continuum elasticity (summarized above) dominates, for smaller beads
however a new, and novel elastic response is found: Consider a
volume $V$ distorted by a force on a particle of size $R$.
Again we take this volume as anisotropic with dimensions $ l \times l
\times D$. 
In this volume the filaments which traverse the
volume bend with a wavelength $l$ and
there is a bending contribution to the total energy, coming from eq.
(\ref{landau}) which varies as
\begin{equation}
\label{e1}
E_1 \sim (l^2 D) (a^2 k_B T l_p/l^4) \rho
\end{equation}
The three multiplicative factors are respectively the volume excited,
the bending energy per unit length of filament and the filament
density within the volume.  $a$ is again the typical amplitude of the
excitation in the volume. To this bending contribution one must add
the equivalent of $E_{var}$. When we impose the bending on the volume
$V$ there is also a variation in the geometry of the confining tubes.
For instance in the direction of $f$ the tubes are compressed by a
factor comparable to $a/D$. Thus there is thus a contribution to the
energy coming from the macroscopic bulk modulus of the form
\begin{equation}
\label{e2}
E_2 \sim (l^2 D) ( (a/l)^2 + (a/D)^2) (\rho k_B T/l_e)
\end{equation}
where we have again respectively the volume, the square elastic stress
and the macroscopic elastic modulus.  We can now optimize $E_1 + E_2$
by minimizing over $D$. However, we first notice that there
are two term linear in $D$ and that depending on the value of $l$ one
or the other will dominate.  If we look at wavelengths
\begin{equation}
\label{lc}
 l< l_c = \sqrt{l_e l_p} \sim \xi^{2/5} l_p^{3/5} 
\end{equation} 
the contribution from $E_1$ dominates over that from $E_2$.  When $l >
l_c$ the second contribution dominates.

We conclude that there is an important new length scale in the
problem.  When we look at excitations with wavelengths greater than
$l_c$ the two contributions in $E_2$ are going to dominate the
elasticity and we are back to the case of normal continuum elastic
theory. However in the short wavelength limit $l < \sqrt{l_e l_p}$ we
find that the elastic energy is given by
\begin{equation}
\label{energy}
E_{eff} \sim a^2 \rho (l^2 D) (k_B T/D^2 l_e + l_p/l^4)
\end{equation}
Minimizing the energy over $D$ we find several surprising results.
Firstly
\begin{equation}
\label{aniso}
 l^2 \sim D \sqrt{l_e l_p} 
\end{equation}
the theorem of St. Vernet does not apply.  Secondly substituting
eq. (\ref{aniso}) in (\ref{energy}) gives
\begin{equation}
\label{efinal}
 E_{eff} \sim \rho a^2k_B T \sqrt{ {{l_p} \over {l_e}}}
\end{equation}
which should be compared with the corresponding result for a normal
solid eq. (\ref{evar2}). From eq. (\ref{aniso}) we see that the volume
excited scales in an anisotropic fashion with the wavelength, quite
unlike normal elastic solids. One is  reminded of the
penetration of excitation into a smectic liquid crystal with an
effective energy for fluctuations of the form \( E \sim K \int[
(\partial_x^2 u)^2 + (\partial_y^2 u)^2 + \beta (\partial_z u)^2 ]  dV \) where
the $z$ axis is defined by the direction of application of the force.

The length scale $l$ is absent in the energy (\ref{efinal}) which has
not been minimized over the wavelength; this is analogous to the case
of a normal elastic material eq.  (\ref{evar2}) in two rather than
three dimensions. It suggests that we are in the lower critical
dimension for the problem (at least in a certain range of wavelengths)
and that a fuller treatment will bring out logarithmic corrections to
our picture.  We also deduce, since eq. (\ref{efinal}) is independent
of $l$, that the amplitude of displacement of a particle should be
independent of its size, in contrast to the dependence discussed above
eq. (\ref{inverse}): We are no longer dominated by the short
wavelength cutoff in the energy integral eq. (\ref{evar}) despite
being in three dimensions

Finally we note that for coherency in this picture we require that the
depth of penetration of the excitation into the sample $D$ be greater
than the distance between the filaments $\xi$, otherwise a continuum
description as used here must break down.  This implies that $l>
l_l=\sqrt{\xi\sqrt{ l_e l_p}} = \xi^{7/10} l_p^{3/10}$.  On
wavelengths shorter than this, one is presumably sensitive to
filaments directly in contact with the probing particle and do not
feel the three dimensional nature of the sample.

Substituting typical values for material constants, $\xi\sim.5\mu$,
$l_p \sim 15\mu$ we find that $l_e \sim 1 \mu$. The crossover length
scale $ l_c \sim 4 \mu$. The short wavelength cutoff $l_l \sim 1.5
\mu$.  We thus expect the following series of crossovers as a function
of probing wavelength. (a) For $\xi< l < l_l$ one probes the bending of
individual filaments. (b) For $l_l < l < l_c$ collective excitations
of the solution become important with anomalous penetration of the
excitation into the sample. (c) For $l> l_c$ the elasticity becomes
isotropic.  These crossovers are too closely spaced to be
experimentally studied in great detail, however we conclude that to
measure a valid macroscopic response function particle sizes should
be substantially greater than $l_c=4\mu$.

Until now we have only considered the low frequency response of a
sample, that is for times long enough that all longitudinal stresses
have relaxed along the tube.  It has been shown \cite{me} that one
expects two plateau moduli as a function of frequency.  The low
frequency plateau used in the above discussion comes from variation in
tube geometry under sample shear. The second much larger contribution
which dominates at higher frequencies comes from coupling of the shear
to the longitudinal density fluctuations of the filament in its tube.
Can we see the crossover between the low frequency and high frequency
behaviour with micro-bead techniques?  This question is difficult to
answer, the static approach used above is not adapted to answer this
dynamic question however we can certainly expect that the frequency of
crossover between the two regimes will vary with the bead size.  

The regime of the high plateau in macroscopic rheology is delimited by
the two times $\tau_e \sim 0.1s$ and $ \tau_e (l_p/l_e)^2 \sim 10s$.
This second time is the time needed for excitations to diffuse a
distance $l_p$ along the tube.  It is important because macroscopic
shear produces density fluctuations along the tube which are coherent
over a distance $l_p$.  When we excite a sample with a wavelength $l$,
which is smaller the $l_p$, we expect that the window of times for the
observation of this high plateau is reduced to the interval between
$\tau_e$ and $\tau_e (l/l_e)^2$. For the smallest beads this
high second plateau should almost completely disappear. Even with
larger beads the elastic modulus should be substantially
underestimated over certain frequency ranges.  More detailed
discussion of this regime seems to be difficult without a detailed
{\it dynamic} theory of the coupling of the bend and longitudinal
degrees of freedom.

To conclude actin mechanics shows a quite rich series of crossover in
the response function $G(q,\omega)$. We have simple arguments for the
wavevector dependence of this function at frequencies between
$10^{-3}Hz$ and $10Hz$.  Further work requires a full dynamic theory
of the coupling between bending and density fluctuations.
$G(q,\omega)$ is more complicated than might be expected; recent
experiments which interpret elastic stress propagation in terms of an
isotropic elastic theory characterized by an elastic constant and
Poisson ratio \cite{sackmannpoisson} may be missing some interesting
physics on length scales smaller than $4\mu$.

\end{document}